\documentclass[sigchi-a, nonacm=true]{acmart}

\usepackage[colorinlistoftodos,prependcaption,textsize=tiny]{todonotes}

\def\BibTeX{{\rm B\kern-.05em{\sc i\kern-.025em b}\kern-.08emT\kern-.1667em\lower.7ex\hbox{E}\kern-.125emX}}
    
\copyrightyear{2019}
\acmYear{2019}
\acmPrice{00.00}

\begin{document}

%
\title{Towards Human-Centered AutoML}

%
\author{Florian Pfisterer, Janek Thomas, Bernd Bischl}
\orcid{0001-8867-762X}
\affiliation{%
  \institution{LMU Munich}
  \city{Munich}
  \country{Germany}
}
\email{florian.pfisterer@stat.uni-muenchen.de}

%
\renewcommand{\shortauthors}{Pfisterer et al.}

%
\begin{abstract}
Building models from data is an integral part of the majority of data science work flows.
While data scientists are often forced to spend the majority of the time available for a 
given project on data cleaning and exploratory analysis, the time available to practitioners
to build actual models from data is often rather short due to time constraints for a given project.
AutoML systems are currently rising in popularity, as they can build powerful models without human oversight. In this position paper, we aim to discuss the impact of the rising popularity of 
such systems and how a user-centered interface for such systems could look like.
More importantly, we also want to point out features that are currently missing
in those systems and start to explore better usability of such systems from a data-scientists perspective.
\end{abstract}

\maketitle

\vspace{-.3cm}
\section{Introduction}
Data Science projects usually involve a multitude of steps to collect useful insights from data.
After general steps of data validation, cleaning and exploration, a data scientist applies preprocessing steps to the data set. 
A large number of possible preprocessing operations are available. 
Features often need to be extracted, scaled, transformed and imputed.
The preprocessed data can then be used to train a machine learning model which can be used to predict new data.
Finding optimal models and preprocessing operations is notoriously difficult as many, often somewhat technical, decisions have to be made and nearly all available operations and algorithms contains additional hyperparameters. 
The idea of automatically obtaining a machine learning pipeline from data is the central element of the research field \textit{Automatic Machine Learning} (AutoML).
AutoML systems are currently rising in popularity as they can find powerful models without human oversight and knowledge.
The aim of this paper is three-fold: We $i)$ describe capabilities of current Auto-ML systems and how they integrate into the data science work-flow, $ii)$ discuss potential shortcomings in current practices and $iii)$ discuss how an interface for the practitioner could ideally look like.
A  model used in production can affect potentially affect millions of user or highly important technical systems or processes.
Hence, such models often have to be judged and optimized w.r.t. to multiple criteria. 
The importance of these criteria, like predictive performance, model size, prediction speed and interpretability will vary between projects and have inherent trade-offs, meaning that not all of them can be optimized equally well.
While data scientists are often forced to spend the majority of the time on data cleaning and exploratory analysis,
the time available to build and investigate actual models from the data is often comparably small (20\% is an often-quoted number).
Automatizing this part, can thus result in better models and reduce mistakes in the process. 
We consider extending current approaches to be able to incorporate multiple criteria a challenge necessary, to significantly advance data-science applications. 



\begin{marginfigure}
  \centering
  \includegraphics[width=\linewidth]{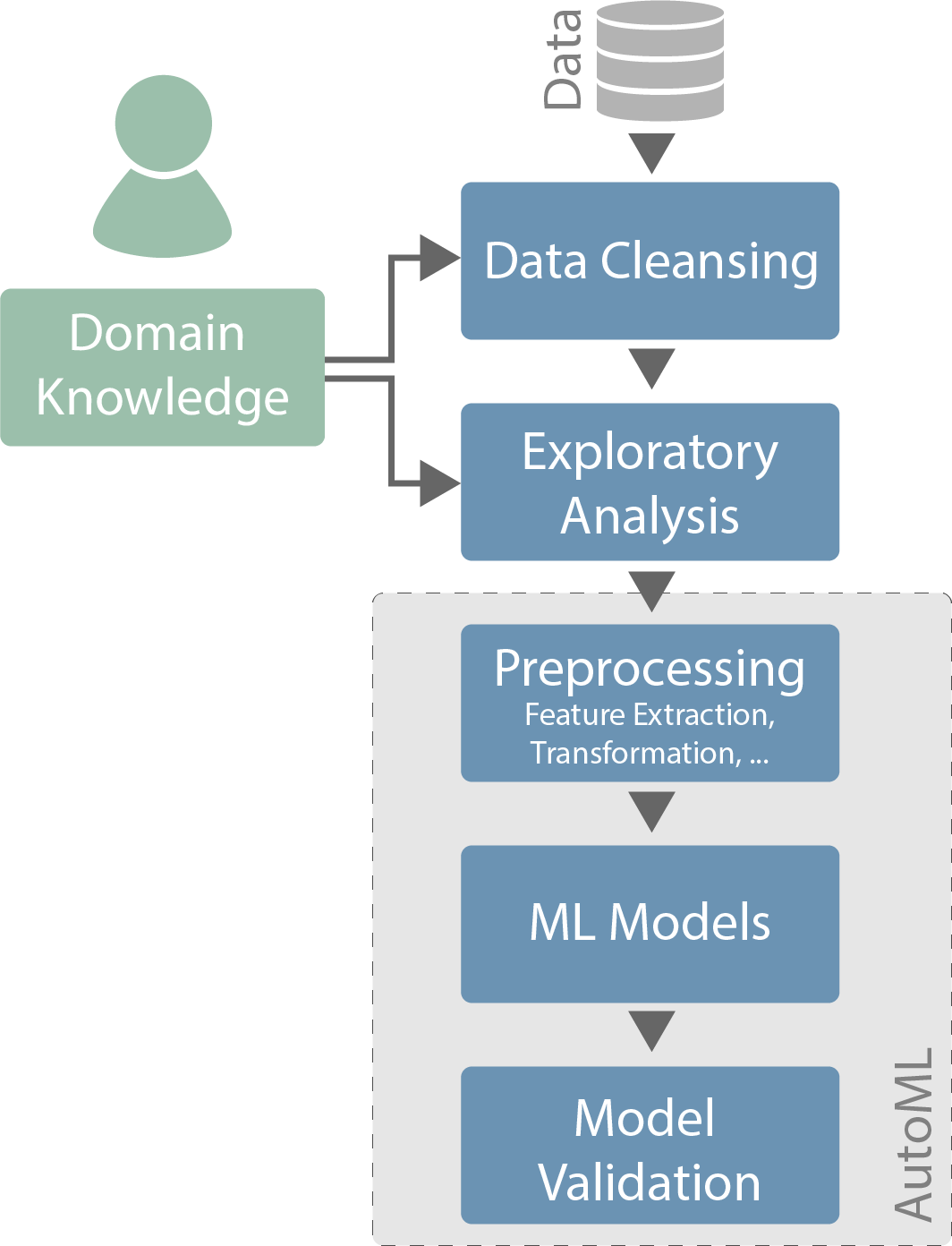}
  \caption{Simplified data analysis workflow}
  \Description{Graph displying the workflow for a typical data science project: AND TODO HUMAN INFLUENCE ON VARIOUS STEPS}
  \label{fig:workflow}
\end{marginfigure}

\section{Status Quo}

In this work, we mainly focus on the machine-learning part of the data science workflow. 
Substantial effort has been put into many components, such as exploratory data analysis and data-cleasning (c.f \textit{the automated statistician project} \cite{ghahramani2015b}), or providing additional visualizations \cite{viz}, but these data science stages still remain often largely manual processes. 
Various machine learning toolboxes are available to users in different programming languages. 
Popular examples are  Weka~\cite{Hall2009} (Java), scikit-learn~\cite{Pedregosa2011} (Python) and mlr~\cite{Bischl2016} (R). 
These toolboxes serve as a first step towards making machine-learning accessible to a wide audience of practitioners and build the foundation of most state-of-the-art AutoML systems. 
While the field of AutoML has obtained a lot of attention in recent years from companies such as Google (Google Cloud AutoML), Amazon (Amazon Sagemaker), it has long been an active field of research and various implementations already exist.
Examples for those include \texttt{auto-weka} \cite{autoweka}, \texttt{auto-sklearn} \cite{autosklearn} and \texttt{tpot} \cite{tpot}.
In the scientific community, those systems are compared in several AutoML challenges organized at top machine-learning conferences \cite{Feurer2018}. 
These challenges focus solely on the predictive performance of models built by the AutoML systems as it is easy to compare and rank the systems in this way.
Other criteria as discussed above are completely ignored and a simpler, sparse model is not preferred to a much more complex one despite having nearly identical predictive performance.


A typical (simplified) workflow involving such a system looks as follows (c.f. figure \ref{fig:workflow}): 
After accessing and cleaning the data, the data scientist conducts exploratory data analysis in order to gather first insights from the data. 
When the data has a sufficient quality the user passes on the data to an AutoML system, which then optimizes a machine learning pipeline of preprocessing, model and hyper-parameters in order to achieve a high predictive performance. 
The quality of the resulting model is then assessed by performance metrics as well as by human domain experts.
Recently, the fields of \textit{fair and transparent machine learning} (FATML) (c.f \cite{fairmlbook}) and \textit{interpretable machine learning} (IML) (c.f. \cite{imlbook}) erupted as important new fields of research. 
Different methods that increase fairness, transparency, and interpretability have been proposed \cite{bigdatablabla, lipton}. 
In many cases, respecting these criteria is crucial for model selection as predictions need to be explained to clients, users or society.


\vspace{-.3cm}
\section{Where is the human in AutoML?}

\begin{marginfigure}
  \centering
  \includegraphics[width=.8\linewidth]{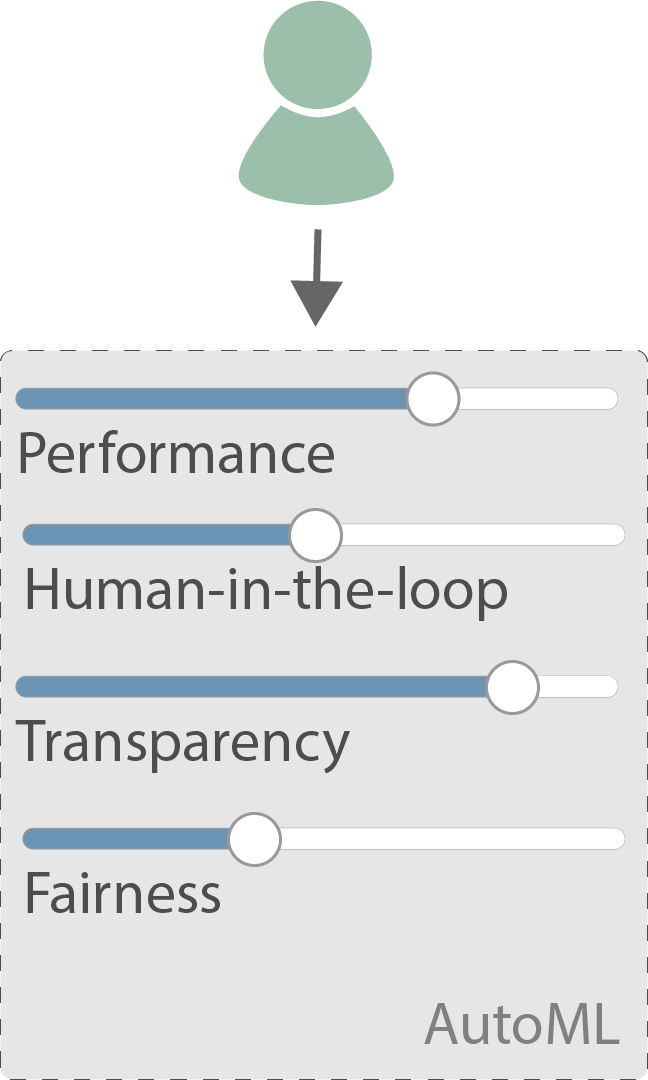}
  \caption{Exemplary parameters for a human-centered AutoML interface}
  \Description{Simplified interface for AutoML Systems}
  \label{fig:interface}
\end{marginfigure} 

The humans role in current AutoML processes is to choose data sets, validation protocols, performance measures to optimize and to define the pipeline search space, i.e., which preprocessing and modeling steps to consider.
After that, the systems does not require human intervention and returns an optimal model after a prespecified amount of time. 
This often drastically speeds up the process of obtaining well working models as technical optimization is left to the machine and has not to be dealt with in a manual trial-and-error process.
Furthermore, this process can be scaled up to run on massively parallel systems nowadays.
A very important approach to making this complex process more accessible to humans was proposed in \cite{atmseer}.
Still, large amounts of time are spent on data-cleaning, preprocessing and hand-crafting features, as these steps typically depend on domain knowledge. 
Their effectiveness can be observed in Kaggle's machine learning competitions, as well as in research \cite{domingos}.
We want to start discussing how humans can be enabled by AutoML systems even further and how those systems need to be extended in order to achieve this. 
A very basic suggestion can be observed in figure \ref{fig:interface}.
We consider the current inability of many AutoML systems to incorporate criteria such as fairness and interpretability a major drawback.
Additionally, systems should make intermediate results available to the practitioner, which can then be evaluated and played back to the AutoML system.
This can especially help in situations, where user preferences are not easily quantifyable, or where relevant criteria are not a-priori known.
The field of AutoML promises great enhancements to the current data science workflow, but to harness its full potential,
it needs to be extended to be more accommodating towards multiple criteria and human intervention. 
%




\noindent This work has been funded by the German Federal Ministry of Education and Research (BMBF) under Grant No. 01IS18036A. The authors of this work take full responsibilities for its content.
\vspace{-.3cm}
\bibliographystyle{ACM-Reference-Format}
\bibliography{main}

\end{document}